\documentclass[aps,nofootinbib,showpacs,showkeys,preprint
,tightenlines ] {revtex4}

\usepackage{epsf,epsfig,subfigure,graphicx,amsmath,amssymb}
\usepackage{color}

\newcommand{\dis}[1]{\begin{equation}\begin{split}#1\end{split}\end{equation}}
\newcommand{\be}{\begin{equation}}
\newcommand{\ee}{\end{equation}}

\newcommand{\eq}[1]{Eq.~(\ref{#1})}
\newcommand{\etal}{et al.}

\newcommand{\Mp}{M_P}

\newcommand{\epstar}{\epsilon_\vp^*}

\newcommand{\ecstar}{\epsilon_\chi^*}

\newcommand{\vp}{\varphi}

\newcommand{\fnl}{f_{\rm NL}}
\newcommand{\fnlf}{f_{\rm NL}}
\newcommand{\tilder}{\tilde{r}}

\def\bkone{{\bf k_1}}
\def\bktwo{{\bf k_2}}

\def\picube{(2\pi)^3}
\newcommand{\sdelta}[1]{\!\delta^{\,3}(\mathbf{#1})}

\newcommand{\calA}{{\cal A}}

\newcommand{\calP}{{\cal P}}

\def\ces{\cos^2\Theta_e}

\def\ses{\sin^2\Theta_e}

\def\bea{\begin{eqnarray}}
\def\eea{\end{eqnarray}}

\begin{document}

 \begin{flushright}
{\tt  APCTP-Pre2011-010
\\PNUTP-11-A03}
\end{flushright}

\title{\Large\bf Natural Hybrid Inflation Model  
\\ with Large Non-Gaussianity}

\author{ Ki-Young Choi$^{(a)}$$^{(b)}$\footnote{email: kiyoung.choi@apctp.org}
and Bumseok Kyae$^{(c)}$\footnote{email: bkyae@pusan.ac.kr} }
\affiliation{$^{(a)}$ Asia Pacific Center for Theoretical Physics, Pohang, Gyeongbuk 790-784, Republic of Korea\\
$^{(b)}$Department of Physics, POSTECH, Pohang, Gyeongbuk 790-784, Republic of Korea\\
$^{(c)}$ Department of Physics, Pusan National University, Busan
609-735, Republic of Korea}


\begin{abstract}
We propose an inflationary model (``natural hybrid model''), which combines the supersymmetric hybrid model and the natural inflation model 
to achieve the spectral index of $0.96$, and the axion decay constant smaller than the Planck scale, $f\ll M_P$. 
By introducing both U(1)$_R$ and a shift symmetry and employing the minimal K$\ddot{\rm a}$hler potential, 
the eta-problem can be still avoided. The two inflaton fields in this model can admit large non-Gaussianity.
\end{abstract}


 \maketitle


\section{Introduction}

Inflation \cite{review} is known to be the best solution to the fine tuning problems associated with the initial conditions of the universe in the standard big bang cosmology: it resolves so-called the ``homogeneity problem'' and ``flatness problem.'' Moreover, 
quantum fluctuation of the inflaton fields generated during inflation can eventually provide seeds
of large scale structure (LSS) observed in our universe~\cite{InflationFluctuation}.
Thus, it has been tested by observing the fluctuations in the LSS, the temperature anisotropy in the cosmic microwave background (CMB), and so on through the satellite experiments such as COBE \cite{COBE} and Wilkinson Microwave Anisotropy Probe (WMAP) \cite{Komatsu:2010fb} so far, and also will be more precisely checked by Planck \cite{Planck} in the near future. 

The recent WMAP seven data \cite{Komatsu:2010fb} shows that the initial density perturbation is almost scale invariant and Gaussian with
 $ n_\zeta =0.96^{+0.014}_{-0.013}$  and $-10 < \fnl^{\rm local}< 74$ at the 95\% confidence level.
The single field chaotic inflation is well consistent with the current observation,
implying very small non-Gaussianity suppressed by the slow-roll parameters~\cite{Maldacena:2002vr}. 
If sizable non-Gaussianity will be detected in Planck satellite experiment in the near future, however, it will play the role of the criterion to select a realistic inflationary model. 

Inflationary scenario is based on scalar field theory. 
Although inflationary scenario seems to be inevitable in cosmology, it is very non-trivial to realize in quantum field theory. In order to keep the small inflaton mass against quantum corrections, introduction of supersymmetry (SUSY) is helpful as in particle physics. Unless an inflationary model is not very elaborately constructed, however, introduction of SUSY is not enough: the positive vacuum energy in supergravity (SUGRA) induces the Hubble scale inflaton mass, yielding $\eta\sim {\cal O}(1)$ during inflation, which destroys the slow-roll condition.  It is called the ``eta-problem.''

``Hybrid inflation'' \cite{Linde:1993cn} is basically an two field model with one as inflaton  and the other, called waterfall field, to terminate the inflation when it becomes tachyonic. The advantage of it is that the inflaton's field value  
is small compared to the Planck scale, and thus it is legitimate to use the low energy effective theory.
In the SUSY version of hybrid inflation \cite{FtermInf2,FtermInf}, the potential can be made  flat enough, avoiding the eta-problem: fortunately the Hubble induced mass term is accidentally cancelled out with the minimal K$\ddot{\rm a}$hler potential and the Polonyi type superpotential during inflation. The specific form of the superpotential can be guaranteed by the introduced U(1)$_R$ symmetry.  

By the logarithmic quantum correction to the scalar potential, the inflaton can be drawn to the true minimum, leading to reheating of the universe by the waterfall fields. Moreover, thanks to such a logarithmic correction, the vacuum expectation values (VEVs) of the waterfall fields can be determined with  the CMB anisotropy \cite{FtermInf}. The VEVs turn out to be tantalizingly close to the scale of the grand unified theory (GUT). Accordingly, the waterfall fields can be regarded as GUT breaking Higgs in this class of models \cite{3221,422,FlippedSU(5),SO(10)}.
This inflationary model predicts a red-tilted power spectrum \cite{FtermInf} around 
\begin{equation}
n_\zeta \approx 1+2\eta\approx 1-\frac{1}{N_f}\approx 0.98
\end{equation}
for $N_f=50-60$ e-folds. It is too large compared to the present bound on the spectral index. 

In the ``natural inflation'' model \cite{Freese:1990rb}, the inflaton field is regarded as a pseudo Nambu-Goldstone boson. 
Hence, a U(1) Peccei-Quinn symmetry, U(1)$_{\rm PQ}$, 
should be assumed to be there, and the inflaton's small mass can be protected against quantum corrections. Its scalar potential, which is given by a sinusoidal functional of the inflaton, can be induced by instanton effects, which break U(1)$_{\rm PQ}$.  
Since intanton effects still respect a shift symmetry,
 the axion does not appear in 
the K$\ddot{\rm a}$hler potential in the SUSY version of the natural inflation model \cite{NaturalChaotic}. As a result, the unwanted Hubble scale inflaton mass term is not induced in the SUGRA potential. 

However, for the slow-roll parameter ``$\eta$'' to be small enough in this model, the Peccei-Quinn breaking scale or the axion decay constant $f$ must be larger than the Planck scale, 
\begin{equation}
f \gtrsim 3 M_P ~, 
\end{equation}
where $M_P$ is the reduced Planck mass ($\approx 2.4\times 10^{18}$ GeV).
It implies that U(1)$_{\rm PQ}$ should be valid above the Planck scale.
However, such U(1)$_{\rm PQ}$ is not natural, because quantum gravity effects are known to break all global symmetries including U(1)$_{\rm PQ}$.\footnote{It might be possible to obtain the effectively large $f$ from sub-Planckian Peccei-Quinn scale though
with multiple axion fields~\cite{Kim:2004rp,Dimopoulos:2005ac}.} 
 
In this Letter, we attempt to improve such shortcomings in the two inflation models, SUSY hybrid and natural inflation models, by combining them. 
That is to say, we will examine the possibility to achieve the spectral index of 0.96 and $f\ll M_{P}$ in the combined model. 
We will call it ``natural hybrid model'' for inflation.
Accordingly, we have two inflaton fields in this model in addition to the waterfall fields.   

One of the important properties of two or multiple fields inflation models is the existence of the
non-adiabatic perturbations during inflation and they can change the evolution of the curvature perturbation after horizon exit \cite{Komatsu:2001rj,Valiviita:2009bp,Hikage:2009rt,Wands:2007bd}. The residual isocurvature perturbation may cause an observably large  non-Gaussianity \cite{Byrnes:2010em}, and becomes adiabatic.
After the isocurvature modes are exhausted, the curvature perturbation is preserved during the radiation era and finally leaves the fossils in the CMB anisotropy observations.

Even though it was  in the slow-roll  phase, the non-adiabatic mode can redistribute the shape of the
perturbation during inflation and finally leads to the non-Gaussian feature at the end of inflation.
The general condition for the large non-Gaussianity was analytically derived
 in \cite{ Choi:2007su,Byrnes:2008wi} with some examples and studied in detail for the multi-field hybrid inflation model in \cite{Byrnes:2008wi2}.
 
There are other multiple field models which generate large non-Gaussianity such as  the curvaton scenario, modulated (p)reheating, and an inhomogeneous
end of inflation.  Interestingly enough, it can be shown that these models have common features in the  mechanism  generating large non-Gaussianity \cite{Alabidi:2010ba}.
The on-going projects to improve the observational sensitivity with Planck satellite \cite{Planck} 
and using LSS data are expected to greatly constrain  
the non-Gaussianity parameters or discover the non-Gaussian feature in the primordial density perturbation in the near future. 
They could discriminate a right one among various mechanisms of density perturbation generation.

This Letter is organized as follows. 
In section II, we will construct an inflationary model and check the SUGRA corrections of the model. 
In section III, we will discuss the various features, particularly the spectral index and large non-Gaussianity predicted in this model.
We will conclude in section IV. 
Finally, section V is devoted to Appendix, in which we presents various expressions related to the spectral index and the non-linear parameter, based on $\delta N$ formalism.

\section{The SUGRA Model}

To preserve the small inflaton masses during inflation against the Hubble scale SUGRA corrections, let us introduce the U(1)$_R$ and a shift symmetry. Under the U(1)$_R$ symmetry, the superpotential $W$ and a superfield $S$ are supposed to undergo the same transformation, i.e. $W\rightarrow e^{2i\gamma}W$ and $S\rightarrow e^{2i\gamma}S$. 
Under the shift symmetry, a superfield $T$ is supposed to transform as $T\rightarrow T+2\pi i f$, where $f$ is a constant with mass dimension one. We also consider the superfields of a conjugate pair, $\psi$ and $\overline{\psi}$, which are assumed to carry opposite gauge charges.  

The superpotential consistent with the U(1)$_R$ and the shift symmetries is written as
\begin{eqnarray} \label{superpot}
W=\kappa S\left[M^2-m^2e^{-T/f}-\psi\overline{\psi}(1+\rho e^{-T/f})\right] ~,
\end{eqnarray}
where $M^2$, $m^2$, $f$, $\kappa$, and $\rho$ are parameters. Actually, $e^{+T/f}$, $e^{\pm 2T/f}$, etc. also can contribute to the superpotential Eq.~(\ref{superpot}).  However, they do not give rise to qualitatively different patterns of inflation, compared to a simple case only with $e^{-T/f}$. So we will neglect them for simplicity.
The ``$\rho$ term'' ($\rho\ll 1$) does not affect the inflationary scenario, since  $\langle\psi\overline{\psi}\rangle=0$ during inflation, as will be discussed later. 
On the other hand, it can be important when $\langle\psi\overline{\psi}\rangle\neq 0$.  
We assume a hierarchy among the dimensionful parameters, $m\ll f\ll M~(\ll M_P)$. It turns out to be necessary for slow-roll of the inflatons and large non-Gaussianity.  
For simplicity, we also assume that they all are real parameters. 

We note that the U(1)$_R$ forbids $S^2$, $S^3$, etc., which destroy slow-roll inflation, and 
restricts the superpotential to the linear form of $S$. This superpotential keeps the small inflaton mass against the large SUGRA correction of the Hubble scale during inflation, as will be seen later. 
The scalar component of $T$ is composed of two real scalar fields, $T(x)=\phi(x)+ia(x)$.\footnote{In this Letter we use the same notation for a superfield and its scalar component.} 
Because of the shift symmetry, the K${\rm \ddot{a}}$helr potential, which is a functional of $T+T^*$, does not contain $a(x)$. Accordingly, the F-term scalar potential in SUGRA is expected not to induce the Hubble scale mass term for $a(x)$ during inflation \cite{NaturalChaotic}.

From Eq.~(\ref{superpot}), we can obtain the F-term scalar potential: 
\begin{eqnarray}\label{V_F}
&&\quad\quad~~ V_F/\kappa^2=\left|M^2-m^2e^{-T/f}-\psi\overline{\psi}(1+\rho e^{-T/f})\right|^2 
\\
&&+|S|^2\left\{(|\psi|^2+|\overline{\psi}|^2)|1+\rho e^{-T/f}|^2+\left|m^2+\rho\psi\overline{\psi}\right|^2\frac{e^{-2\phi/f}}{f^2}\right\} ~. \nonumber
\end{eqnarray}
The SUSY minimum is located at $S=a=0$ and $M^2-m^2e^{-T/f}-\psi\overline{\psi}(1+\rho e^{-T/f})=0$. The D-term potential constrains the VEVs of $\psi$ and $\overline{\psi}$ to satisfy $|\psi|=|\overline{\psi}|$. We assume that they develop VEVs in the real direction: $\langle\psi\rangle=\langle\psi_r\rangle/\sqrt{2}
=\langle\overline{\psi}\rangle=
\langle\overline{\psi}_r\rangle/\sqrt{2}$, where the subscript ``$r$'' indicates the real component for each field. 
At the minimum of Eq.~(\ref{V_F}), then, the VEV of $a$ vanishes. Since we have only the two constraints in the field space $(\psi,\overline{\psi},\phi)$, i.e. $M^2-m^2e^{-\phi/f}-\psi_r\overline{\psi}_r(1+\rho e^{-\phi/f})/2=0$ and $\psi_r=\overline{\psi}_r$, the VEVs of $\phi$ and $\psi_r\overline{\psi}_r$ are not fixed yet, and so there potentially remains a modulus. However, by including the soft SUSY breaking mass terms for them, 
$\delta V_{\rm soft}=(m_\phi^2\phi^2+m_\psi^2\psi_r^2
+m_{\overline{\psi}}^2\overline{\psi}_r^2)/2$, 
their VEVs can be determined,
\begin{eqnarray} \label{phi}
&&\frac{\phi}{f} ~e^{\phi/f}\approx-2\frac{m_\psi^2}{m_\phi^2}\frac{m^2}{f^2} \sim {\cal O}\left(m^2/f^2\right)~\ll~1 ~, \\
&&\psi_r^2\approx 2(M^2-m^2e^{-\phi/f} - \kappa^{-2}m_\psi^2)\approx 2(M^2-m^2-\kappa^{-2}m_\psi^2) ~,
\end{eqnarray}      
where $m_\phi$, $m_\psi$, and $m_{\overline{\psi}}$ indicate the soft mass parameters for the real components of $\phi$, $\psi$, and $\overline{\psi}$. They are of the TeV scale. We set  $m_\psi=m_{\overline{\psi}}$ for simplicity. From Eq.~(\ref{phi}), 
we see $\phi/f\approx -2(m_\psi^2 m^2)/(m_\phi^2 f^2)\sim {\cal O}( m^2/f^2)\ll 1$. Around the minimum of the scalar potential, the mass eigenstates $\{\Phi,\Psi\}$ and their corresponding mass squared turn out to be 
\begin{eqnarray}
\Phi\approx\phi+\epsilon\psi_r  ~,~~m_\Phi^2\approx m_\phi^2 ~;~~~{\rm and}~~~
\Psi\approx\psi_r-\epsilon\phi~,
~~m_\Psi^2\approx 4\kappa^2 M^2~,
\end{eqnarray}
where $\epsilon\equiv m^2e^{-\phi/f}/\sqrt{2}fM$. Thus, the VEVs of $\Phi$ and $\Psi$ are 
\begin{eqnarray}
\langle\Phi\rangle\approx
\langle\phi+\epsilon\psi_r\rangle\sim{\cal O}\left(m^2/f\right)
~,~~~{\rm and}~~~\langle\Psi\rangle\approx
\langle\psi_r-\epsilon\phi\rangle\approx \sqrt{2}M ~.
\end{eqnarray}
We will focus on the case $m^2/f\ll 10^{10}$ GeV, because this case turns out to yield large non-Gaussianity. 

To obtain a sufficient inflation, we suppose $S\gtrsim M$ initially. Since $|S|^2$ plays the role of the mass squareds for $\psi$ and $\overline{\psi}$, the initial condition $S\gtrsim M$ compels the VEVs of $\psi$ and $\overline{\psi}$ to vanish during inflation. 
As a result, $S$ and $T$ can be light enough and the scalar potential becomes dominated by the positive vacuum energy density $\kappa^2 M^4$, only if the mass of $S$ in Eq.~(\ref{V_F}) is much lighter than the Hubble scale, i.e. $\kappa^2 m^4/f^2\ll \kappa^2 M^4/M_P^2$ or $m^4/M^4\ll f^2/M_P^2~(\ll 1)$.   
Thus, the condition $S\gtrsim M$ provides a quasi-flat scalar potential with positive vacuum energy density. Under the condition, thus, inflation can arise.

The positive vacuum energy density during inflation breaks SUSY explicitly. The coupling between $S$ and the waterfall fields $\psi$, $\overline{\psi}$ in Eq.~(\ref{superpot}) and mass splittings between the scalars and fermions by SUSY breaking  
induce the quantum correction at one loop to the scalar potential Eq.~(\ref{V_F}):
\begin{eqnarray} \label{deltaV}
\delta V_F\approx \mu^4\times\frac{\kappa^2}{8\pi^2}~{\rm log}\frac{S}{\Lambda} ~,
\end{eqnarray}  
where $\mu^4$ ($\equiv \kappa^2 M^4$) denotes the vacuum energy density during inflation and $\Lambda$ the renormalization scale \cite{FtermInf}. It can draw the inflaton $S$ to the true minimum of $S=0$. Even with this logarithmic functional of $S$, the slow roll conditions still holds. We assume it is dominant over the mass term of $S$ proportional to $m^4/f^2$ in Eq.~(\ref{V_F}). 

As $S$ approaches to $M$, $\psi$ and $\overline{\psi}$ become destabilized, since they become tachyonic for $S<M$ around the origin. Thus, $\psi$ and $\overline{\psi}$
start rolling down to their true minima discussed in Eq.~(\ref{V_F}), when $S$ becomes smaller than $M$. It means that the VEVs $\langle \psi\rangle$ and $\langle\overline{\psi}\rangle$, which play the role of the masses for $S$ as seen in Eq.~(\ref{V_F}), grow from zero approximately up to $M$. We note that before $\psi$ and $\overline{\psi}$ reach $M$, the slow-roll condition for $S$ breaks down when $\eta_s\equiv M_P^2\partial_S^2V/V\approx 1$, that is to say, 
\begin{eqnarray} \label{endinf}
M_P^2~\kappa^2\left(\langle |\psi|^2\rangle +\langle |\overline{\psi}|^2\rangle\right)\approx
 V ~,
\end{eqnarray} 
where $V$ is approximately given by Eq.~(\ref{V_F}).  
This is {\it the condition for end of inflation} in this model. 
Since the leading term of $V$ is a large constant ($\kappa^2M^4$), the left-hand side of Eq.~(\ref{endinf}) should be so. For instance,  $\langle\psi\rangle$ ($=\langle\overline{\psi}\rangle$) is approximately given by $M^2/M_P\approx 10^{13}$ GeV, if $M\approx 5\times 10^{15}$ GeV.  

With the D-flat condition, $|\psi|=|\overline{\psi}|=|\psi_r|/\sqrt{2}$, Eq.~(\ref{endinf}) is written in terms of $\psi_r$,   $M_P^2\kappa^2\psi_r^2\approx\kappa^2[{\cal M}^4
-({\cal M}^2-|S|^2)\psi_r^2+\frac{\psi_r^4}{4}]$, where ${\cal M}^4\approx M^4(1-2\frac{m^2}{M^2}{\rm cos}\frac{a}{f})$, and we neglect the term proportional to $m^4$. 
Since the waterfall field $\psi_r$ is sitting always at the local minimum, its VEV is determined by $\partial_{\psi_r} V=0$, yielding  $\frac{\psi_r^2}{2}={\cal M}^2-|S|^2=\sqrt{{\cal M}^4-V/\kappa^2}$. 
It recasts Eq.~(\ref{endinf}) into $V^2+4\kappa^2M_P^4V-4\kappa^4M_P^4{\cal M}^4=0$ or $V=2\kappa^2M_P^4(-1+\sqrt{1+{\cal M}^4/M_P^4})\approx \kappa^2M^4$, i.e. $V(S,a)$ is almost a constant ($\approx \kappa^2M^4$) when inflation is over. It implies that the condition for end of inflation almost respects a uniform energy density condition in this case. Hence, one can expect that the end point effect in  non-Gaussianity would be quite small \cite{CKK}.\footnote{
In Ref.~\cite{Naruko}, the end point effects in non-Gaussianity, particularly for the cases that two inflaton masses are degenerate ($m_1^2=m_2^2$) and  hierarchical ($m_1^2\gg m_2^2$),  have been  studied. 
Following the notation of the Ref.~\cite{Naruko},  $G=2\kappa^2|S|^2$ and $\sigma^2=2\kappa^2{\cal M}^2$, respectively, in our case.
Thus, our model corresponds to the limit, $g_1=\alpha=\delta=0$, $\beta=\gamma=\pi/2$ of  Ref.~\cite{Naruko},  and $m_1^2$ and $m_2^2$  are estimated as $\eta_a|_{S=M}$ and $\eta_s|_{S=M}$, respectively, and as will be seen  later, $\eta_a|_{S=M}\gg\eta_s|_{S=M}$ in our case. 
With such limits of the parameters, 
the condition for end of inflation corresponds to a  straight line in the field space, and
 the end point effect in non-Gaussianity can not be large as noticed in~\cite{Huang:2009vk,CKK}. It is consistent with the analysis of Ref.~\cite{Naruko}.}
    
Once the slow-roll condition for $S$ violated, that of $a$ is also violated quickly.   
During $M^2-\psi\overline{\psi}\gg m^2+\rho\psi\overline{\psi}$ in Eq.~(\ref{V_F}), the slow-roll parameter of axion $\eta_a$ ($\equiv M_P^2\partial_a^2V/V$) can be smaller than the unity even with $f\ll M_P$. As $M^2-\psi\overline{\psi}$ decreases and becomes comparable with $m^2$ or $\rho\psi\overline{\psi}$, 
however, the slow-roll conditions for $a$ becomes also violated, since $\eta_a$ becomes $M_P^2/f^2$ ($\gg 1$) in this case. Thus, $a$ should also roll down to zero.
Due to the existence of the non-adiabatic mode, the total curvature perturbation evolves until the end of inflation and abruptly stops to evolve after slow-roll is violated.
The perturbation of waterfall fields during the waterfall dynamics does not change the curvature perturbation in the large scales as studied recently in many literatures~\cite{ Lyth:2010ch,Abolhasani:2010kr,Fonseca:2010nk,Gong:2010zf,Lyth:2010zq,Abolhasani:2011yp}.

Finally, let us check SUGRA corrections. We ignore the quantum correction Eq.~(\ref{deltaV}) for a while.  
By including SUGRA corrections, the Hubble scale mass terms for some scalar fields could be induced during inflation. To see which fields acquire such masses, let us consider the full F-term scalar potential in SUGRA Lagrangian. Since $\psi$ and $\overline{\psi}$ got already heavy masses proportional to $S$ even in global SUSY, we will not consider them.  

During inflation, thus, the superpotential and the  K${\rm \ddot{a}}$hler potential are given by
\begin{eqnarray}
W=\kappa S\left(M^2-m^2e^{-T/f}\right) ~,~~~K=|S|^2+\frac14(T+T^*)^2 ~,
\end{eqnarray}
respectively. 
They respect the U(1)$_R$ and shift symmetries. Here, the K${\rm \ddot{a}}$hler potential takes the minimal form without containing $T-T^*$. {\it The resulting  kinetic terms of the scalar fields are of the canonical type}. We will see no Hubble induced mass term for $S$ appears in the scalar potential \cite{SUGRAcorr}. If there exists a quartic term of $S$ in the K${\rm \ddot{a}}$hler potential, however, it would generate the unwanted Hubble induced mass term, which destroys the slow-roll conditions for $S$. Hence, we should assume its coefficient is small enough ($\lesssim 10^{-2}$), if it exits. This assumption needs to be justified by a proper UV theory of SUGRA for its naturalness in the future.  Since the shift symmetry is encoded in the K${\rm \ddot{a}}$hler potential, only the real part of $T$, i.e. $\phi(x)$ [$=(T+T^*)/2$] appears in it. Due to the reason, we will see that the Hubble induced mass term for $a(x)$ [$=(T-T^*)/2i$] does not appear in the SUGRA potential. 

With the covariant derivatives in SUGRA, 
\begin{eqnarray}
&&D_SW=\frac{\partial W}{\partial S}+\frac{W}{M_P^2}\frac{\partial K}{\partial S}=\mu^2\left(1+\frac{|S|^2}{M_P^2}\right)\left(1-\frac{m^2}{M^2}e^{-T/f}\right) ~, \\
&&D_TW=\frac{\partial W}{\partial T}+\frac{W}{M_P^2}\frac{\partial K}{\partial T}\approx \mu^2\frac{fS}{M_P^2}\left(\frac{\phi}{f}+\frac{M_P^2}{f^2}\frac{m^2}{M^2}e^{-T/f}\right) ~, 
\end{eqnarray}
one can write down the F-term scalar potential:
\begin{eqnarray} \label{V_sugra}
&& V_{\rm SUGRA}=e^{K/M_P^2}\left[\left|D_SW\right|^2+2\left|D_TW\right|^2-3\frac{|W|^2}{M_P^2}\right] \nonumber \\
&&\approx\mu^4\left(1+\frac{\phi^2}{M_P^2}\right)\left[\left|1-\frac{m^2}{M^2}e^{-T/f}\right|^2+2\frac{f^2|S|^2}{M_P^4}\left|\frac{\phi}{f}+\frac{M_P^2}{f^2}\frac{m^2}{M^2}e^{-T/f}\right|^2\right] \\
&& \approx \mu^4\left[1-2\frac{m^2}{M^2}e^{-\phi/f}{\rm cos}\frac{a}{f} +\frac{\phi^2}{M_P^2}\right] ~,  \nonumber 
\end{eqnarray}
where we used $e^{K/M_P^2}\approx (1+\phi^2/M_P^2)(1+|S|^2/M_P^2)$, and $K^{SS^*}=1/K_{SS^*}=1$, $K^{TT^*}=1/K_{TT^*}=2$, $K_{ST^*}=0$, etc. We dropped $|S|^4/M_P^4$ in Eq.~(\ref{V_sugra}), because of its smallness.
Due to the presence of the last term, $\mu^	4\phi^2/M_P^2$ ($\approx 3H^2\phi^2$) in the last line of Eq.~(\ref{V_sugra}), which spoils the slow-roll condition for $\phi$, 
$\phi$ should be stabilized somewhere during inflation. 
On the other hand, the quadratic term of $S$ was  cancelled out, and so it does not appear in Eq.~(\ref{V_sugra}). It is a nice feature of the SUSY hybrid inflation model \cite{FtermInf2}. Since ``$e^{K/M_P^2}$'' in the scalar potential does not contains $a$, the Hubble induced mass term of it also does not appear in Eq.~(\ref{V_sugra}). It results from the shift symmetry \cite{NaturalChaotic}. 

In our study we require the hierarchy between the parameters
\begin{equation} \label{hierarchy}
\frac{m^2}{M^2}~\ll~\frac{f^2}{M_P^2} ~\ll~1 ~.
\end{equation}
We will see later that this condition is necessary for slow-roll of $a$.
Note that $f$ is regarded as being smaller than $M_P$.
In this case, the VEV of $\phi$ is estimated as $|\phi/f|\approx{\cal O}(M_P^2m^2/f^2M^2)\ll 1$. So we can neglect ``$e^{-\phi/f}$'' and ``$\phi^2/f^2$'' in Eq.~(\ref{V_sugra}).

\section{Natural Hybrid Inflation}

We are particularly interested in the following form of the potential for inflation with two scalar fields: 
\begin{eqnarray} \label{inf}
V_{\rm inf}=\mu^4\left(1+\alpha {\rm log}\frac{S}{\Lambda}-\lambda {\rm cos}\frac{a}{f}\right) ~,
\end{eqnarray}
where the scalar fields $S(x)$ and $a(x)$ play the role of the inflatons.
It can be derived from Eq.~(\ref{V_F}) or (\ref{V_sugra}) by taking $\psi=\overline{\psi}=0$,  $\phi/f\ll 1$, and Eq.~(\ref{hierarchy}).
The dominant vacuum energy comes from the first term $\mu^4$.
The second term in Eq.~(\ref{inf}) is originated from the scenario of the SUSY hybrid inflation  and third term from the natural inflation. As discussed in Eq.~(\ref{deltaV}), the second term is generated by SUSY breaking effects at one loop \cite{FtermInf} and  so $\alpha =\kappa^2/8\pi^2$. We assume a hierarchy between the dimensionless parameters $\alpha$ and $\lambda$, $1\gg\alpha\gg\lambda$. 
Thus, the dynamics of the trajectory in the field space is mostly dominated by $S$ field, though the cosmological observables such as the spectral index $n_\zeta$ and non-linear parameter $f_{\rm NL}$ are controlled by both fields.
By comparing Eq.~(\ref{inf}) with Eq.~(\ref{V_sugra}), the parameters $\mu^4$ and $\lambda$ can be identified as 
\begin{equation} \label{lambda}
\mu^4=\kappa^2M^4~,\quad \textrm{and}\quad  \lambda\equiv \frac{2m^2}{M^2}e^{-\phi/f}\approx \frac{2m^2}{M^2} ~.
\end{equation}  

With Eq.~(\ref{inf}) the slow-roll parameters for $S$ and $a$ are estimated as 
\begin{eqnarray}
\epsilon_s=\frac{\alpha^2M_P^2}{2S^2}\equiv\frac{\alpha}{2\chi^2} ~,&& ~~~~\eta_s=-\frac{\alpha M_P^2}{S^2}\equiv-\frac{1}{\chi^2} ~, \label{Slow-Roll}\\
\epsilon_a=\frac{M_P^2\lambda^2}{2f^2}~{\rm sin}^2\frac{a}{f}\equiv \frac{\xi^2\lambda}{2}~{\rm sin}^2\theta~, &&~~~~\eta_a=\frac{M_P^2\lambda}{f^2}~{\rm cos}\frac{a}{f}\equiv \xi^2{\rm cos}\theta ~, \nonumber
\end{eqnarray}
where the fields and parameters were simplified as
\begin{eqnarray} \label{simplepara}
\chi\equiv \frac{S}{\sqrt{\alpha}M_P}~,~~ \theta\equiv \frac{a}{f}~,~~~ {\rm and}~~~ \xi^2\equiv \frac{M_P^2}{f^2}\lambda ~. 
\end{eqnarray}
Hence, the slow-roll conditions ($\epsilon_{s,a}\ll 1$, $|\eta_{s,a}|\ll 1$) are fulfilled as long as $\alpha$ and $\lambda$ 
are small enough and
\begin{eqnarray}
\chi^2\gg 1~,~~{\rm and}~~\xi^2=\frac{M_P^2}{f^2}\lambda ~\ll ~1 ~,
\end{eqnarray}  
where we see that the second condition is consistent with our previous requirement Eq.~(\ref{hierarchy}).
Note that if $\lambda$ is quite small, ``$f$'' can be much smaller than $M_P$  unlike in the original natural inflation scenario.

During the slow roll inflation the equations of motion for $S$ and $a$ fields 
are simplified to be
\dis{
&3H\dot{S} + \frac{\alpha \mu^4}{S} + 4\kappa^2\psi^2 S =0\\
&3H\dot{\theta} +\frac{\lambda \mu^4}{f^2}\sin\theta =0.\label{EOM}
}
Neglecting the $\psi $ term, which vanishes during inflation, they give the solutions  for the $S$ (or $\chi$) and $\theta$ ($=a/f$), 
\begin{eqnarray} \label{efolds}
\chi_*^2-\chi_e^2=2N_f ~,~~~{\rm and}~~~{\rm tan}\frac{\theta_*}{2}={\rm tan}\frac{\theta_e}{2}~ e^{\xi^2N_f} ~, \label{EOM_sol}
\end{eqnarray}
where $N_f$ is the e-folding number between the horizon exit and the end of inflation,
which is assumed to be around $N_f \approx 50-60$.
Since we require slow-roll condition until the end of inflation, we take $\chi_*^2>2N_f\sim 100$.
Throughout this Letter, the subscripts (or superscripts) ``$*$'' and ``$e$'' denote the  values evaluated at a few Hubble times after horizon exit  and the end of inflation, respectively.   Since inflation is over when $S_e\approx M$ ($=\mu/\sqrt{\kappa}$), we have the relation, $V|_e\approx\mu^4\approx 8\pi^2\alpha^3M_P^4\chi_e^4$.  

Since we are interested in the evolution of the curvature perturbation in the superhorizon scale,
we use $\delta N$ 
formalism~\cite{starob85,ss1,Sasaki:1998ug,lms,lr}. The number of e-foldings, $N$, is given by 
\dis{
 N=\int^{t_{\rm{e}}}_{t_*} H(t)dt,
}
and the curvature perturbation is evaluated as
\dis{
\zeta=\delta N  = \sum_I
N_{,I}\delta\vp_{I*}+\frac12\sum_{IJ}N_{,IJ}\delta\vp_{I*}\delta\vp_{J*}+\cdots\,, 
}
where $N,_I=\partial N/\partial \vp^I_*$ and the index $I$ runs over all of the inflaton fields.
We will consider the
power spectrum and bispectrum defined (in the momentum  space) by
\begin{eqnarray}\label{powerspectrumdefn} \langle\zeta_{\bkone}\zeta_{\bktwo}\rangle &\equiv&
\picube\,
\sdelta{\bkone+\bktwo}\frac{2\pi^2}{k_1^3}\calP_{\zeta}(k_1) \, , \\
\langle\zeta_{{\mathbf k_1}}\,\zeta_{{\mathbf k_2}}\,
\zeta_{{\mathbf k_3}}\rangle &\equiv& \picube\, \sdelta{{\mathbf
k_1}+{\mathbf k_2}+{\mathbf k_3}} B_\zeta( k_1,k_2,k_3) \,. \end{eqnarray}
With the $\delta N$ formalism the power spectrum, the spectral index and the non-linear parameter are given by 
\begin{eqnarray}
{\cal P}_\zeta&=&\sum_I N_{,I}^2 {\cal P}_*,\qquad {\cal P}_*\equiv \frac{H_*^2}{4\pi^2}\label{spectrum} \\
n_{\zeta}-1&=& -2\epsilon^* +
\frac{2}{H}\frac{\sum_{IJ}\dot{\vp}_J N_{,JI}N_{,I}}{\sum_K N_{,K}^2},\label{index}
\\
\fnl&=&\frac56 \frac{\sum_{IJ}N_{,IJ}N_{,I}N_{,J}}{\left(\sum_I N_{,I}^2\right)^2},
\end{eqnarray}
if the inflatons' kinetic terms are given by the canonical form.
Especially for the separable potential we can calculate e-folding number analytically
and so the analytic expression of the corresponding $\fnl$ can be obtained~\cite{Vernizzi:2006ve,Choi:2007su}.
For sum potentials, we summarize them in the Appendix.
Therefore, the curvature perturbation evolves after horizon exit during the inflation and the non-Gaussianity can be developed, which can be observed through the satellite experiments in the near future~\cite{Byrnes:2008wi,Byrnes:2008wi2}.  More examples of large non-Gaussianity generated during the inflation with the adiabatic limit were studied recently~\cite{Kim:2010ud,Gong:2011cd,Elliston:2011dr, Mulryne:2011ni}.

Field fluctuations are generated at the horizon exit with almost Gaussian statistics, and thus the curvature perturbation is also Gaussian at that time. 
However, due to the non-adiabatic modes in the multiple field inflation, the curvature perturbation  
experiences non-linear evolution and changes its shape to be non-Gaussian. 
When the non-Gaussianity becomes larger than $\mathcal O(1)$ during inflation, we can approximate it analytically as in the papers~\cite{Byrnes:2008wi,Byrnes:2008wi2}.
 The size of the non-linear parameter $\fnl$ is determined by the slow-roll parameters  $\epsilon_i$'s and $\eta_i$'s. Especially the sign of $\fnl$ is given by the relative size of $\eta$ between the horizon exit and the end of inflation by $-\eta_* + 2\eta_e$ of a relevant field~\cite{Byrnes:2008wi}.

\begin{figure}[!t]
\begin{center}
   \includegraphics[width=0.5\textwidth]{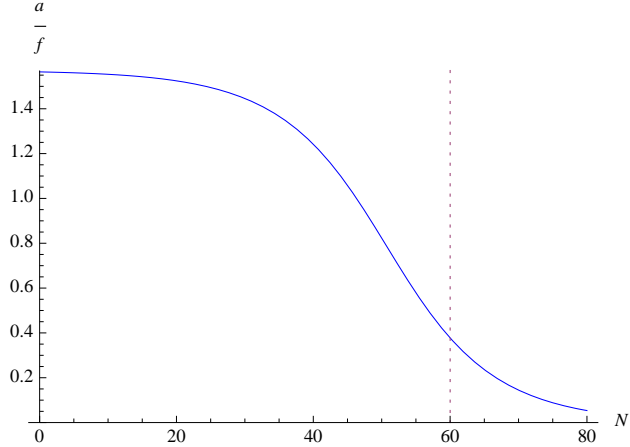}
\end{center}
\caption{Evolution of axion field without waterfall field (with blue line).  However, the inflation ends at e-folding number $N_f=60$ (with vertical dotted line) and the axion decouples from the blue line and moves quickly to the minimum of the potential.}
\label{fig:evol_A}
\end{figure}

\begin{figure}[!t]
\begin{center}
   \includegraphics[width=0.5\textwidth]{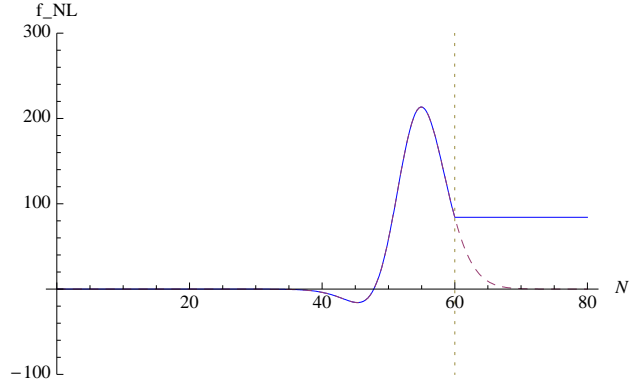}
\end{center}
\caption{Evolution of $\fnl$ for case of a positive $\fnl$ (Case 1 in Table~\ref{tb:models}). The dashed line corresponds to the case without waterfall fields, and the solid line is for the model with the waterfall fields, where end of inflation occurs at $N_f=60$ (with dotted line).}
\label{fig:evol_fnl1}
\end{figure}

In our model, we have two inflaton fields $S$ and $a$, which are slowly running during inflation with the waterfall fields, $\psi$ and $\overline{\psi}$ fixed at the origin. Large non-Gaussianity can arise when the axion field $a$ is located around the top
at the horizon exit. Since the axion field is on the ridge where the potential is concave, i.e. $\eta_a <0$, the trajectories diverge and 
large negative non-Gaussianity is generated. 
After the axion field crosses the convex point, $a/f=\pi/2$, the potential changes the curvature and
$\eta_a>0$, which makes the change of the sign of $\fnl$.\footnote{The behaviors of $\fnl$ on the ridges and valleys are also well described in the paper of Elliston \etal~\cite{Elliston:2011dr}.  See also Ref.~\cite{hilltopNonG}.} 
 Soon the waterfall field is destabilized and the inflation get to the end, when the slow-roll condition is violated. However, the waterfall dynamics does not affect the curvature perturbation after slow-roll inflation as studied in Refs.~\cite{Lyth:2010ch,Abolhasani:2010kr,Fonseca:2010nk,Gong:2010zf,Lyth:2010zq,Abolhasani:2011yp}.
The dynamics of the axion field and the evolution of $\fnl$ are shown in Figs.~\ref{fig:evol_A} and \ref{fig:evol_fnl1} respectively. Here we used the parameters of Case 1 in Table~\ref{tb:models}. 
 
In the region of large non-Gaussianity, we can make good estimates for the power spectrum and non-Gaussianity using the analytic study in the separable potential in Refs.~\cite{Byrnes:2008wi,Byrnes:2008wi2}. In this region of large $|\fnl|$ with $\epsilon_s\gg\epsilon_a$, we get\begin{eqnarray}
{\rm SIN}^2\Theta\equiv\frac{\epsilon_a}{\epsilon_s+\epsilon_a}\approx\left(\frac{\lambda}{\alpha}\right)\xi^2\chi^2~{\rm sin}^2\theta\ll 1 ~,
\end{eqnarray}
where we used \eq{Slow-Roll}.
Therefore the power spectrum (${\cal P}_\zeta$), spectral index ($n_\zeta$),  and tensor-to-scalar ratio ($r$) are approximately given in terms of the slow-roll parameters: 
\begin{eqnarray} \label{Psp}
&&{\cal P}_\zeta \approx\frac{V_*}{24\pi^2M_P^4\epsilon_s^*}(1+\tilde{r}) ~, \\
&&n_\zeta -1 \approx -2\epsilon_s^*+2\frac{-2\epsilon_s^*+\eta^*_s+\tilde{r}\eta^*_a}{1+\tilde{r}} ~, \label{SI}\\
&&r\approx \frac{16\epsilon_s^*}{1+\tilde{r}} ~, 
\end{eqnarray}
where $\tilde{r}$ is defined as the ratio of the contribution to the curvature perturbation from each field by
\begin{eqnarray} \label{rtilde}
\tilde{r}\equiv \frac{N_{,a}}{N_{,S}} \approx\frac{{\rm SIN}^4\Theta_e}{{\rm SIN}^2\Theta_*}\approx\xi^2\left(\frac{\lambda}{\alpha}\right)\frac{\chi_e^4~{\rm sin}^4\theta_e}{\chi_*^2~{\rm sin}^2\theta_*} ~.
\end{eqnarray} 
In Appendix, the relevant original expressions are summarized. 

%

\begin{figure}[!t]
\begin{center}
   \includegraphics[width=0.5\textwidth]{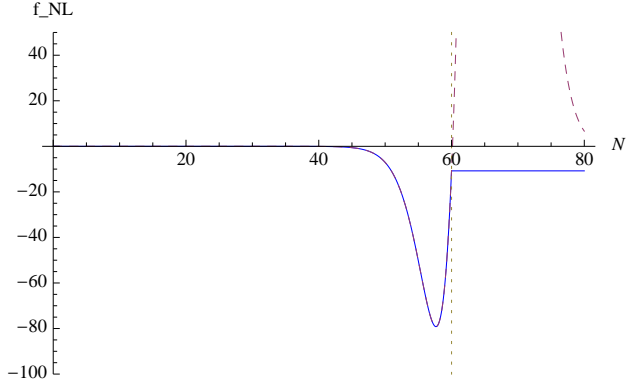}
\end{center}
\caption{Evolution of $\fnl$ for case of a negative $\fnl$. The dashed line is without waterfall fields and the solid line is for the model with the waterfall fields, where end of inflation occurs at $N_f=60$ (with dotted line).}
\label{fig:evol_fnl4}
\end{figure}

In the limit of $\tilde{r}\rightarrow 0$, the linear term of the curvature perturbation is dominated by $S$ field and the power spectrum and the spectral index is determined solely by the $S$ field perturbation.   
Thus it reproduces the result of the SUSY hybrid inflation, i.e. $n_\zeta\approx 1+ 2\eta^*_s \approx 0.98$  from \eq{Slow-Roll}  and \eq{SI} with $\chi^2_*>2N_f$. 
In the opposite limit, for large $\tilde{r} \gg 1$, the power spectrum comes dominantly from the $a$ field perturbation and, the spectral index is
 \dis{
 n_\zeta -1 \approx 2\eta_a^*=2\xi^2\cos\frac{a_*}{f} ~,  
 }
so that we can chose $\xi^2\approx 0.02$ to make $n_\zeta\approx 0.96$ for $a_*/f\sim \pi$.
To get large positive $\fnl$, however, we need $\xi^2\gtrsim 0.1$, because we need  $\theta_e < \pi/3$
from  $\fnl\propto -\eta_a^*+2\eta_a^e\propto -\cos\theta_*+2\cos\theta_e\approx -1+2\cos\theta_e$, 
and equation of motion  \eq{EOM_sol} 
\dis{
\tan\frac{\theta_e}{2} \approx \frac{2~e^{-\xi^2 N_f}}{\pi-\theta_*}.
}
This predicts 
\dis{
n_\zeta \lesssim 0.8 ~,
}
which is not consistent with WMAP results.
Therefore, the only available value for $\tilder$ compatible with large positive $\fnl$ and the spectral index 
is $\tilde{r}={\mathcal O}(1)$.  For small $\fnl$, however, there should be no problem even with much  smaller $\xi^2$.
We listed some of cases which can predict large non-Gaussianity in Table~\ref{tb:models}.
In Fig~\ref{fig:evol_fnl4}, we show the evolution of $\fnl$ for a case with negative non-Gaussianity.

\begin{table}
\begin{tabular}{c||ccccc|cc|cc}
\hline  
Case&~$\mu~(=\sqrt{\kappa}M)$& ~$\alpha~(=\kappa^2/8\pi^2)$ &~ $\lambda~(\approx 2m^2/M^2)$ &  $\qquad f \quad\qquad$ &  $\pi-\theta_*$ & $\xi^2$&$\chi_*$ & $\tilde{r}$&$\fnl$
\\
\hline \hline
1&$1.7\cdot 10^{14}$ & $2.6\cdot 10^{-8}$ & $5.0\cdot 10^{-13}$ & $5.4\cdot 10^{12} $ & $1.3\cdot 10^{-2}$ &$0.10$ & $15.9$ &$0.19$   &43
\\
\hline 
2&$1.7\cdot 10^{13}$ & $8.6\cdot 10^{-11}$ & $2.4\cdot 10^{-17}$ & $3.8\cdot 10^{10}$  & $1.5\cdot 10^{-2}$ &$0.10$ & $86.1$ &$0.24$   &72
\\
\hline 
3&$9.3\cdot 10^{11}$ & $7.8\cdot 10^{-14}$ & $2.0\cdot 10^{-22}$ & $1.1\cdot 10^{8}$  & $1.5\cdot 10^{-2}$ &$0.096$ & $887$ &$0.23$   &67
\\
\hline 
\end{tabular}
\caption{Some parameter values which give large $\fnl$ at the end of inflation with $N_f=60$. The values of $\mu$ and $f$ are listed in the unit of GeV.  Here we imposed the constraints from Power spectrum \eq{wmapP} and the spectral index \eq{wmapdata}.  
$M$ ($=\mu/\sqrt{\kappa}$) and $f$ should be around the GUT and intermediate scales, respectively. 
}
\label{tb:models}
\end{table}

In Table~\ref{tb:models}, we imposed 
the data from seven years of WMAP~\cite{Komatsu:2010fb},
\begin{eqnarray}
&&{\cal P}_\zeta=2.43\pm 0.115\times 10^{-9}~, \label{wmapP}\\
&&n_\zeta=0.96^{+0.014}_{-0.013} ~~~({\rm assuming}~~r=0) ~ \label{wmapdata}.
\end{eqnarray}
By comparing Eqs.~(\ref{Psp}) and $(\ref{SI})$ with Eqs.~(\ref{wmapP}) and (\ref{wmapdata}),
we can get the relations
\begin{eqnarray} \label{PS}
&&\quad\quad\quad \frac23  \alpha^2\chi_e^4\chi_*^2(1+\tilde{r})\approx 2.43\times 10^{-9} ~,\\ \label{n_s}
&&\frac12\alpha (3+\tilde{r})\approx -1+\tilde{r}\chi_*^2\xi^2{\rm cos}\theta_*+0.02(1+\tilde{r})\chi_*^2 ~. 
\label{nzeta}
\end{eqnarray}
Hence, $\alpha$ and $\tilde{r}$ can be determined in terms of $\chi_*$ (or $\chi_e$), $\theta_*$ , and $\xi^2$. 
Since $\alpha$ is a quite small positive number,  neglecting the left-hand side of \eq{nzeta}, 
$\tilde{r}$ would be
\dis{
\tilde{r}\approx\frac{-1/\chi_*^2+0.02}{\xi^2-0.02}~ ,
 }
 for $ \chi_*\gg 1$ and $\theta_*\approx\pi$. Once $\tilde{r}$ is determined, 
$(\lambda/\alpha)$ is also done by Eq.~(\ref{rtilde}).
We note that $n_s\approx 0.96$ is possible in the presence of the two inflaton fields $S$ and $a$, unlike in the original SUSY hybrid inflation model.

The tensor-to-scalar ratio is suppressed by the slow-roll parameter $\epsilon_s^*$ which magnitude is  smaller than $\eta_s^*$ by a factor $\alpha$. Therefore in this model the tensor-to-scalar ratio is negligible.

If the non-Gaussianity is large, namely for ${\rm SIN}^2\Theta_*\ll {\rm SIN}^2\Theta_e\ll 1$, it is approximately estimated as \cite{Byrnes:2008wi}
\begin{eqnarray}
f_{\rm NL}\approx\frac{5~\tilde{r}^2}{6~{\rm SIN}^2\Theta_e ~(1+\tilde{r})^2}\left(-\eta^*_a+2\eta^e_a\right)
\approx\frac{5~\tilde{r}^2}{6(\lambda/\alpha)\chi_e^2 ~{\rm sin}^2\theta_e(1+\tilde{r})^2}\left(-{\rm cos}\theta_*+2{\rm cos}\theta_e\right)~.
\end{eqnarray}
Since $\chi_e$, $\theta_e$, $\lambda/\alpha$, $\alpha$ and $\tilde{r}$ can be given in terms of $\chi_*$, $\theta_*$, and $\xi^2$ by Eqs.~(\ref{efolds}),  (\ref{rtilde}), (\ref{PS}) and (\ref{n_s}), 
$f_{\rm NL}$ can be determined only if $\chi_*$, $\theta_*$, and $\xi^2$ are given.\footnote{Once $\chi_e$, $\alpha$, $\lambda$ are known, then the SUGRA parameters $\kappa$, $M^2$, $m^2$, and $f$ can be determined via $\alpha=\kappa^2/8\pi^2$, $\chi_e^2= S_e^2/\alpha M_P^2\approx M^2/\alpha M_P^2$, $\lambda\approx 2m^2/M^2$, and $\xi^2=M_P^2\lambda/f^2$ from Eqs.~(\ref{lambda}),  and (\ref{simplepara}).}
The condition for large non-Gaussianity $|f_{\rm NL}|\gtrsim 1$ reads \cite{Byrnes:2008wi}
\begin{eqnarray}
\tilde{r}\left(\frac{1}{\sqrt{{\rm SIN}^2\Theta_e{\cal G}_p}}-1\right)\gtrsim 1 ~,~~~{\rm where}~~~ {\cal G}_p=\frac65 \left|-\eta^*_a+2\eta^e_a\right|^{-1}\approx \frac{6}{5 ~\xi^2|-{\rm cos}\theta_*+2{\rm cos}\theta_e|} ~.~~~
\end{eqnarray}
From this condition, the following three necessary conditions can be derived \cite{Byrnes:2008wi}:
\begin{eqnarray}
{\rm SIN}^2\Theta_* < \frac{3^4}{3\cdot 4^4}\frac{1}{{\cal G}_p^2} ~~&{\rm or}&~~\left(\frac{\lambda}{\alpha}\right)\frac{\chi_*^2 ~{\rm sin}^2\theta_*}{\xi^2(-{\rm cos}\theta_*+2{\rm cos}\theta_e)^2}< 0.07 ~, \\
{\rm SIN}^2\Theta_e < \frac{1}{{\cal G}_p}~~&{\rm or}&~~\left(\frac{\lambda}{\alpha}\right)\frac{\chi_e^2 ~{\rm sin}^2\theta_e}{|-{\rm cos}\theta_*+2{\rm cos}\theta_e|}<0.83 ~, \\
\frac{{\rm SIN}^2\Theta_e}{{\rm SIN}^2\Theta_*}>4{\cal G}_p ~~&{\rm or}&~~\frac{\chi_*^2{\rm sin}^2\theta_*}{\chi_e^2{\rm sin}^2\theta_e}~\frac{1}{\xi^2|-{\rm cos}\theta_*+2{\rm cos}\theta_e|}<0.21 ~.
\end{eqnarray}
They are useful to search for the conditions for large non-Gaussianity. 
We see that {\it large non-Gaussianity is possible, only if $\lambda/\alpha$ and $\pi-\theta_*$ are small enough}. 
For $\theta_*\approx \pi$, thus, a positive large $f_{\rm NL}$ requires $\theta_e<2\pi/3$. 
However, we don't discuss here how the inflaton $a$ is initially set on the top of the potential ($\theta_*\approx\pi$), since it is beyond the scope of this Letter. 
For the maximum and minimum values of $f_{\rm NL}$ in the 95$\%$ confidence level of the WMAP 7-years data   \cite{Komatsu:2010fb}, 
\begin{equation}
-10 < f_{\rm NL}^{\rm local} < 74 ~~~(95\% ~~{\rm CL}) ~.
\end{equation}
%

\begin{figure}[!t]
\begin{center}
   \includegraphics[width=0.5\textwidth]{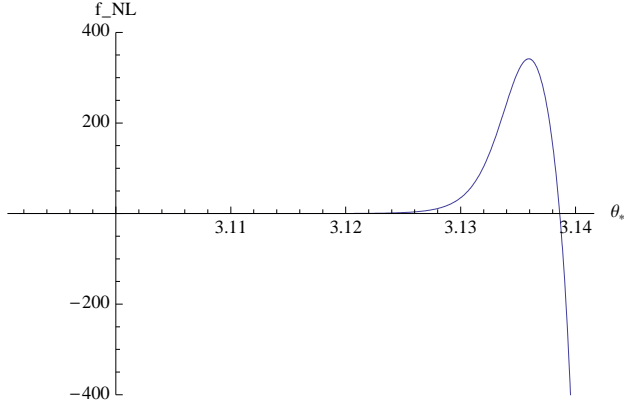}
\end{center}
\caption{Variation of $\fnl$ for different initial positions of the axion field with the same parameters used in Case 1.}
\label{fig:fnl_theta}
\end{figure}

%
When the axion-like inflaton field $a$ is around the top of the potential, i.e. $\theta\approx\pi$, quantum fluctuation of the field would become important and much affects its motion. 
For dominance of the classical motion, the required condition on the background trajectory is $|\dot{a}|\pi/H^2> \sqrt{3/2}$~\cite{Creminelli:2008es,Byrnes:2008wi2}.
This condition constrains the initial value of $a$ ($=f\theta_*$). Using \eq{EOM}, it can be recast as  
\dis{
\sin\theta_* > \frac{f\mu^2}{\sqrt2 \pi \lambda\Mp^3}~. 
}
The initial values listed in Table~\ref{tb:models} fulfill the condition. 

For large $f_{\rm NL}$, 
$M$ ($=\mu/\sqrt{\kappa}$) and $f$ should be around the GUT and intermediate scales as seen in Table~\ref{tb:models}. As discussed earlier, $M$ is the VEV scale of $\psi$, $\overline{\psi}$ after inflation over, and $f$ 
could be the PQ symmetry breaking scale. Thus, the reasonable range of the parameters can yield the values required in particle physics.

In Fig.~\ref{fig:fnl_theta}, we show $\fnl$ at the end of inflation with different initial values of the axion. The large positive $\fnl$ is possible for the axion located around the top, ($\theta_*\lesssim \pi$).  However, it becomes negative if the axion is so close to the top.

\section{Conclusion}

In this Letter, we have constructed an inflationary model by combining the conventional SUSY hybrid and natural inflation models to achieve $n_\zeta\approx 0.96$ and $f\ll M_P$. U(1)$_R$ and a shift symmetry together with the minimal K${\rm\ddot{a}}$hler potential are essential to avoid the eta-problem.  However, the SUSY hybrid inflaton sector needs to make a dominant contribution to the vacuum energy density during inflation. The non-adiabatic mode from the dynamics of the two canonical fields during inflation are responsible for large non-Gaussianity and small tensor-to-scalar ratio. The symmetry breaking scale by the waterfall fields and $f$ in this model should be around the grand unification and intermediate scales, respectively, for large non-Gaussianity.

\acknowledgments{ \noindent  This work was supported by the 
National Research Foundation of Korea (NRF) grant funded by the 
Korea government (MEST) (No. 2010-0009021 and No. 2011-0011083).
K.Y.C. acknowledges the Max Planck Society (MPG), the Korea Ministry of
Education, Science and Technology (MEST), Gyeongsangbuk-Do and Pohang
City for the support of the Independent Junior Research Group at the Asia Pacific
Center for Theoretical Physics (APCTP).
}


\section{Appendix}

$W(\vp,\chi)=U(\vp) + V(\chi)$.
Defining
\begin{eqnarray}
u \equiv \frac{U_*+Z_e}{W_*} ~, \quad \quad
v \equiv \frac{V_*-Z_e}{W_*} ~, \label{uvsum}
\end{eqnarray}
with
\begin{eqnarray}
Z_e &=& \frac{(V_e {\epsilon^e_\vp} - U_e
{\epsilon^e_\chi})}{\epsilon^e}=V_e\ces -U_e\ses ~, \label{Z2}
\end{eqnarray}
the power spectrum and spectral index are given by~\cite{Vernizzi:2006ve}:
\dis{
{\cal P}_\zeta = \frac{W_*}{24\pi^2\Mp^4 }\left(\frac{u^2}{\epstar}
+ \frac{v^2}{\ecstar}\right) ~,\label{spectrum_s}
}
\dis{
n_\zeta-1=-2\epsilon^* -4\frac{u\left(1-\frac{\eta_{\vp}^*}{2\epstar}u\right)
+ v\left(1-\frac{\eta_{\chi}^*}{2\ecstar}v\right)}{u^2/\epstar + v^2/\ecstar} ~.\label{index_s}
}
The non-linear parameter $\fnlf$ is~\cite{Vernizzi:2006ve}:
\begin{eqnarray}
 f_{\rm NL}= \frac{5}{6}
\frac{2}{\left( \frac{u^2}{\epsilon_\vp^*}
+ \frac{v^2}{\epsilon_\chi^*} \right)^2}
\left[
\frac{u^2}{\epsilon^*_\vp}
\left(1  - \frac{\eta^*_{\vp}}{2 \epsilon^*_\vp} u \right)
+ \frac{v^2}{\epsilon^*_\chi}
\left(1
 - \frac{\eta^*_{\chi}}{2 \epsilon^*_\chi } v \right)
+ \left( \frac{u}{{\epsilon^*_\vp}}
- \frac{v}{{\epsilon^*_\chi}} \right)^2 \calA_S
\right] ~,
\label{fNLsum}
\end{eqnarray}
where we define
\begin{eqnarray}
\hat\eta &\equiv&
\frac{(\epsilon_\chi\eta_{\vp}+\epsilon_\vp\eta_{\chi})}{\epsilon}
=\eta_{\vp}\sin^2\Theta+\eta_{\chi}\cos^2\Theta ~, \\
\calA_S &\equiv& - \frac{W_e^2}{W_*^2} \frac{\epsilon^e_\vp
\epsilon^e_\chi}{(\epsilon^e)^2} \left[\epsilon^e -\hat\eta^e \right] =  - \frac{W_e^2}{W_*^2}
\ces\ses \left[\epsilon^e -\hat\eta^e \right]
\label{AS} ~.
\end{eqnarray}

The slow-roll parameters are
\dis{
\epsilon_\vp
=\frac{\Mp^2}{2}\left(\frac{\partial_{\varphi}U}{U+V}\right)^2=\epsilon \cos^2\Theta ~,\qquad
\epsilon_\chi
=\frac{\Mp^2}{2}\left(\frac{\partial_{\chi}V}{U+V}\right)^2=\epsilon \sin^2\Theta ~,
}
and
\dis{
\eta_{\vp}=\Mp^2\frac{\partial_\vp^2 V}{U+V} ~,\qquad
\eta_{\vp\chi}=\partial_{\vp\chi}^2V=0 ~,\qquad
\eta_{\chi}=\Mp^2\frac{\partial_\chi^2 V}{U+V} ~.
}

\end{document}